\documentclass[12pt]{elsart}
\usepackage[cp1250]{inputenc}
\usepackage[IL2]{fontenc}
\usepackage[english]{babel}

\usepackage{amsmath}
\usepackage{graphicx}
\usepackage{xcolor}

\newcommand{\Cerenkov}{$\rm \check{C}$erenkov $\,$}

\begin{document}
\begin{frontmatter}
\title{Fundamentals of cathodoluminescence in a STEM: The impact of sample geometry and electron beam energy on light emission of semiconductors}
\author[USTEM]{Michael St\"oger-Pollach}\footnote{corresponding author: e-mail: stoeger@ustem.tuwien.ac.at, Tel.: +43 (0)1 58801 45204, Fax: +43 (0)1 58801 9 45204}
\author[USTEM,CEITEC,TUB]{Kristýna Bukvi$\rm\check{s}$ová}
\author[USTEM]{Sabine Schwarz}
\author[CEITEC,TUB]{Michal Kvapil}
\author[CEITEC,TUB]{Tomá$\rm\check{s}$ $\rm\check{S}$amo$\rm\check{r}$il}
\author[TUB]{Michal Horák}

\address[USTEM]{University Service Centre for Transmission Electron Microscopy (USTEM), Technische Universit\"at Wien, Wiedner Hauptstra\ss e 8-10, 1040 Wien, Austria}
\address[CEITEC]{Central European Institute of Technology (CEITEC), Brno University of Technology, Purky$\check{n}$ova 123, Brno 612 00, Czech Republic}
\address[TUB]{Institute of Physical Engineering, Brno University of Technology, Technická 2, Brno 616 69, Czech Republic}

\date{\today}

\begin{abstract}
Cathodoluminescence has attracted interest in scanning transmission electron microscopy since the advent of commercial available detection systems with high efficiency, like the Gatan Vulcan or the Attolight Mönch system. In this work we discuss light emission caused by high-energy electron beams when traversing a semiconducting specimen. We find that it is impossible to directly interpret the spectrum of the emitted light to the inter-band transitions excited by the electron beam, because the \Cerenkov effect and the related light guiding modes as well as transition radiation is altering the spectra. Total inner reflection and subsequent interference effects are changing the spectral shape dependent on the sample shape and geometry, sample thickness, and beam energy, respectively. A detailed study on these parameters is given using silicon and GaAs as test materials.

\end{abstract}

\end{frontmatter}
\section{Introduction}
Cathodoluminescence (CL) is a well known technique for many years and often used in the scanning electron microscope (SEM). In recent years it attracts more interest also for scanning transmission electron microscopy (STEM), because two manufacturers (Gatan and Attolight) are offering excellent detectors and spectrometers with very high efficiency. The most attractive use of these kind of detectors at present is plasmonics at metallic nano-objects but also semiconductor research. The spectra recorded in plasmonics can be easily interpreted. The peak maxima are related to plasmonic eigenmodes of the investigated objects \cite{kociak2014a}. In semiconductor research many more difficulties arise when trying to interpret CL spectra.\\

The understanding of optical properties of semiconducting materials at the nanometer scale opens the door for the design of new electronic and opto-electronic devices or functionalized materials\cite{goetsch2018PRM}\cite{goetsch2017ECSt}\cite{miller2017SensAcB}. A possible technique for analyzing optical properties is valence electron energy loss spectrometry (VEELS) \cite{stoeger2008micron}, but it suffers from the dominant elastic peak covering the weak interband transition signal. Another possibility is employing CL, where photons emitted from the recombination process of a former electron-hole-pair excitation is detected \cite{kociak2014b}. The emission of light during the recombination process is further on called "incoherent CL". Unluckily some other physical light emitting processes are superimposing the incoherent CL signal. These are the excitation of transition radiation \cite{frank1966SovPhysUsp}\cite{stoeger2017um_a} and the emission of \Cerenkov radiation \cite{cerenkov1934DAN}\cite{stoeger2006micron}\cite{horak2015um}.

Yamamoto an co-workers \cite{yamamoto2001}\cite{yamamoto1996JElecMic} have already done a detailed study on the light emission of semiconductors under electron beam irradiation. It includes the first hint about the influence of the sample thickness on the emitted spectrum, but a systematic study with respect to sample thickness, beam energy and sample geometry can not be found anywhere in literature up to now. Beside this influence of the sample thickness also the beam energy and the presence of interfaces or edges are altering the emitted light spectrum. This has a tremendous impact on the interpretation of CL spectra from semiconducting materials. In the present study we compare experimental data with simulations using Yamamoto's theory of light emission for slab geometry\cite{yamamoto2001}. Additionally we study the influence of the lateral dimensions of a sample and the presence of sample edges. This is leading to a completely new picture of CL spectra of semiconductors recorded by employing a (scanning) transmission electron microscope (S/TEM).\\

Consequently we will define the CL signal throughout the whole manuscript as any light being emitted by the specimen: incoherent CL, \Cerenkov light and transition radiation, respectively.

\section{Theory}

As long as the \Cerenkov-effect is not excited, a particle uniformly moving in a homogenous medium radiates no electromagnetic waves. To initiate radiation a heterogeneity must be created in the medium trough which the particle passes. The simplest case of an heterogeneity is a planar interface between two homogenous media. The theory for the emission of radiation at such an infinite planar boundary (and two boundaries) was derived in \cite{termikaelian1972} and then thoroughly analyzed by Yamamoto for example in \cite{yamamoto2001}. Such an interface causes light reflection and refraction being well described by the well-known formulae of Fresnel. The analog phenomenon an electron passing through an infinite planar boundary with normal incidence from medium 1 to medium 2, was examined by V.L. Ginzburg and I.M. Frank \cite{ginzburg1946}. The photon flux of transition radiation can be written as

\begin{gather}
\dfrac{d ^2 N}{d \lambda d \Omega}=\frac{\alpha \beta ^2}{\pi \lambda}n_2 \sin^2 \theta _2 \cos^2 \theta_2 \times \nonumber\\
\times\left| \frac{(\epsilon_1-\epsilon_2)(1-\beta^2\epsilon_2-\beta\sqrt{\epsilon_1-\epsilon_2 \sin^2 \theta_2})}{(1-\beta^2\epsilon_2\cos^2\theta_2)(1-\beta\sqrt{\epsilon_1-\epsilon_2\sin^2\theta_2})(\epsilon_1\cos\theta_2+\sqrt{\epsilon_1\epsilon_2-\epsilon_2^2\sin^2\theta_2})}\right| ^2 =\\   
=\frac{\alpha \beta ^2}{4
\pi \lambda}n_2 \sin^2 \theta_2 \left| A(\epsilon_1,\epsilon_2,\theta_2,\beta) \right|^2,\nonumber
\label{TR_equation}
\end{gather}

where $\alpha$ is the Sommerfeld's constant ($\approx \frac{1}{137}$), $\beta=v/c$, $\epsilon_1 $ and $ \epsilon_2$ are the complex dielectric functions of media 1 and 2, respectively, $n_2$ is the refractive index of medium 2, and $\theta_2$ is the emission angle measured from the boundary normal in the direction $\vec{v}$. 
\newline

\begin{figure}[h!]
\begin{center}
\includegraphics[width=10cm]{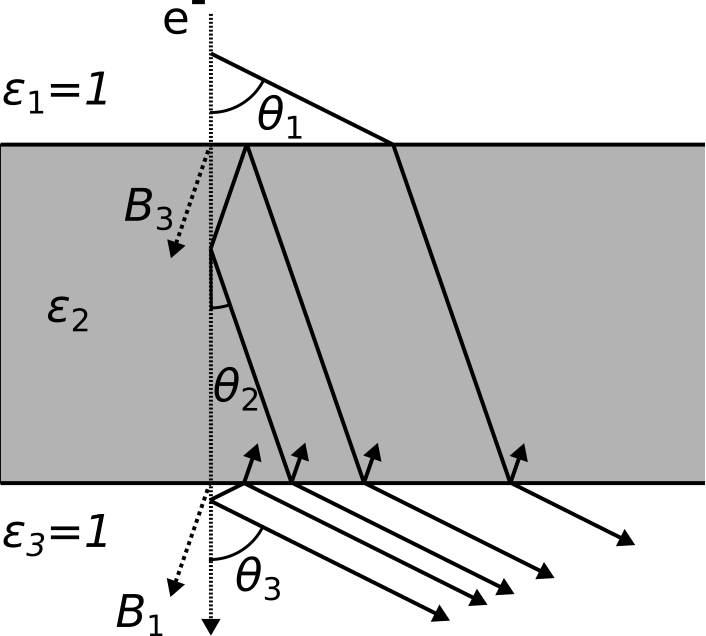}
\end{center}
\caption{Forward photon emission generated by an electron passing through a thin film with dielectric function $\epsilon_2$. (Adopted from \cite{yamamoto1996JElecMic}). The rays $B_1$ and $B_3$ are expressed in Eq. \ref{Dd}.}
\label{schema}
\end{figure}

For transition radiation at two boundaries (as shown in Fig.~\ref{schema}), where medium 1 and 3 is vacuum, the formula of the photon flux for forward emission can be expressed as \cite{yamamoto2001}

\begin{equation}
\dfrac{d ^2 N}{d \lambda d \Omega}=\frac{\alpha \beta ^2}{4\pi^2 \lambda}\sin^2 \theta _3  \left| B_1+B_2\exp(2i\delta) + B_3 \exp(i(\delta-\eta)) \right|^2,
\label{dN}
\end{equation}
where \begin{gather}
B_1=A(\epsilon_2,1,\theta_3,\beta)\nonumber\\
B_2=-\frac{1}{n_2}r_{21} f_{32}D(d)A(1,\epsilon_2,\theta_2,-\beta) \nonumber\\
B_3=\frac{1}{n_2}f_{32}D(d)A(1,\epsilon_2,\theta_2,\beta)  \nonumber
\end{gather}
\begin{equation}
D(d)=(1-r_{21}^2\exp(2i\delta))^{-1},\textrm{ }\delta=\frac{2\pi d n_2 \cos \theta_2}{\lambda},\textrm{ }\eta=\frac{2\pi d}{\lambda\beta} \nonumber
\label{Dd}
\end{equation}
\begin{equation}
f_{ik}=\frac{2n_i\cos\theta_i}{n_k\cos \theta_i+n_i\cos\theta_k},\textrm{ } r_{ik}=\frac{n_k\cos\theta_i-n_i\cos\theta_k}{n_k\cos\theta_i+n_i\cos\theta_k},\nonumber
\end{equation}
\newline
where the values of $\theta_1$ and $\theta_3$ are equal and $d$ is the thickness of medium $\epsilon_2$. The angles are bound by Snell's law and therefore
\begin{equation}
\sin{\theta_1}=n_2\sin{\theta_2}.
\end{equation}

\Cerenkov radiation is included in term $B_2$. To obtain the formula for backward emission, equation~(\ref{dN}) must be modified by replacing the subscripts 1,~2,~3 by 3,~2,~1 and changing $\beta$ to $-\beta$. The angles $\theta$ are then measured from the direction $-\vec{v}$. For normal incidence, the light is p-polarized.
\newline

\section{Experimental}

In order to test the influence of sample geometry, thickness and beam energy on the emitted spectrum, we used undoped silicon. Silicon is one of the best studied materials in semiconductor science and therefore the ideal candidate for this study. \\

\begin{figure}[h!]
\begin{center}
\includegraphics[width=10cm]{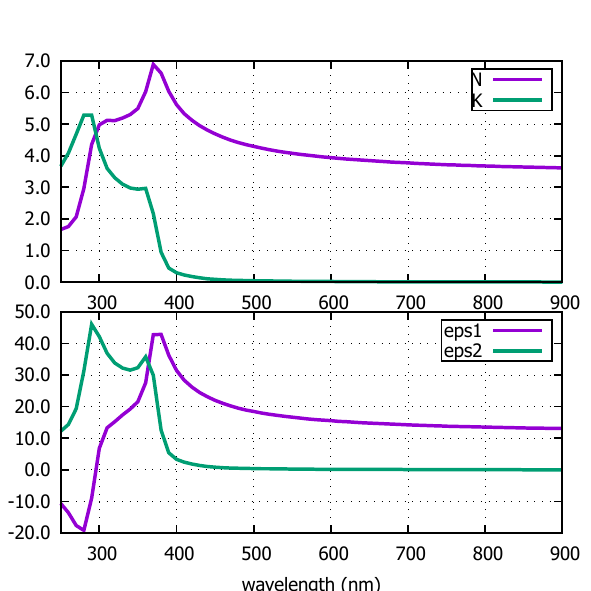}
\end{center}
\caption{Top: Values of $N,$ $K$ for silicon as functions of the wavelength. The refractive index $n_{Si}=N+iK$. Bottom: Values of real and imaginary part of dielectric function of silicon. $\epsilon=\epsilon_1+i\epsilon_2$.}
\label{nk}
\end{figure}

Fig.~\ref{nk} shows real and imaginary part of refractive index of silicon in the interval $\lambda\in\left\langle  250,800\right\rangle $ nm. The values taken from \cite{green2008SEMSC} are a tabulation of multiple data available in literature. The refractive index $n$ is found to be influenced by surface roughness and possible surface layers of oxide but the differences within the various data sets are becoming more significant for photon energies higher than those detected by the system used in this work.

The experiments were performed on a FEI TECNAI transmission electron microscope. The high tension series was recorded in TEM-mode, whereas all other experiments were performed in STEM mode at 200 keV beam energy. For the light detection a GATAN VULCAN system was applied, which allows for recording the emitted light using only the upper or the lower mirror as well as using both mirrors simultaneously. All CL spectra are corrected for the reflectivity of the highly polished Al-mirrors, the absorbency of the light guides and the detection quantum efficiency of the CCD-detector, as well as for the blazing of the 150 lines/mm grating. For the correction the spectrum of the transition radiation of Aluminum was recorded, compared to theory, and finally the ratio between experiment and theory was applied as correction function for all CL spectra \cite{stoeger2017um_a}.

\section{Results}
In order to get a complete picture of light emission under electron beam irradiation, we have done five experiments. First, we varied the beam energy keeping the sample thickness constant. Second, we varied the sample thickness leaving the beam energy constant. Third, we varied the distance of measurement from the edge of the sample at a constant beam energy and constant sample thickness. Fourth, we varied the sample geometry by fabricating small silicon ribbons with constant thickness but various widths. Fifth, we investigated an interface between nano-crystalline silicon (nc-Si) and crystalline silicon (c-Si) at constant beam energy.

\subsection{varying high tension}

\begin{figure}[h!]
\begin{center}
\includegraphics[width=10cm]{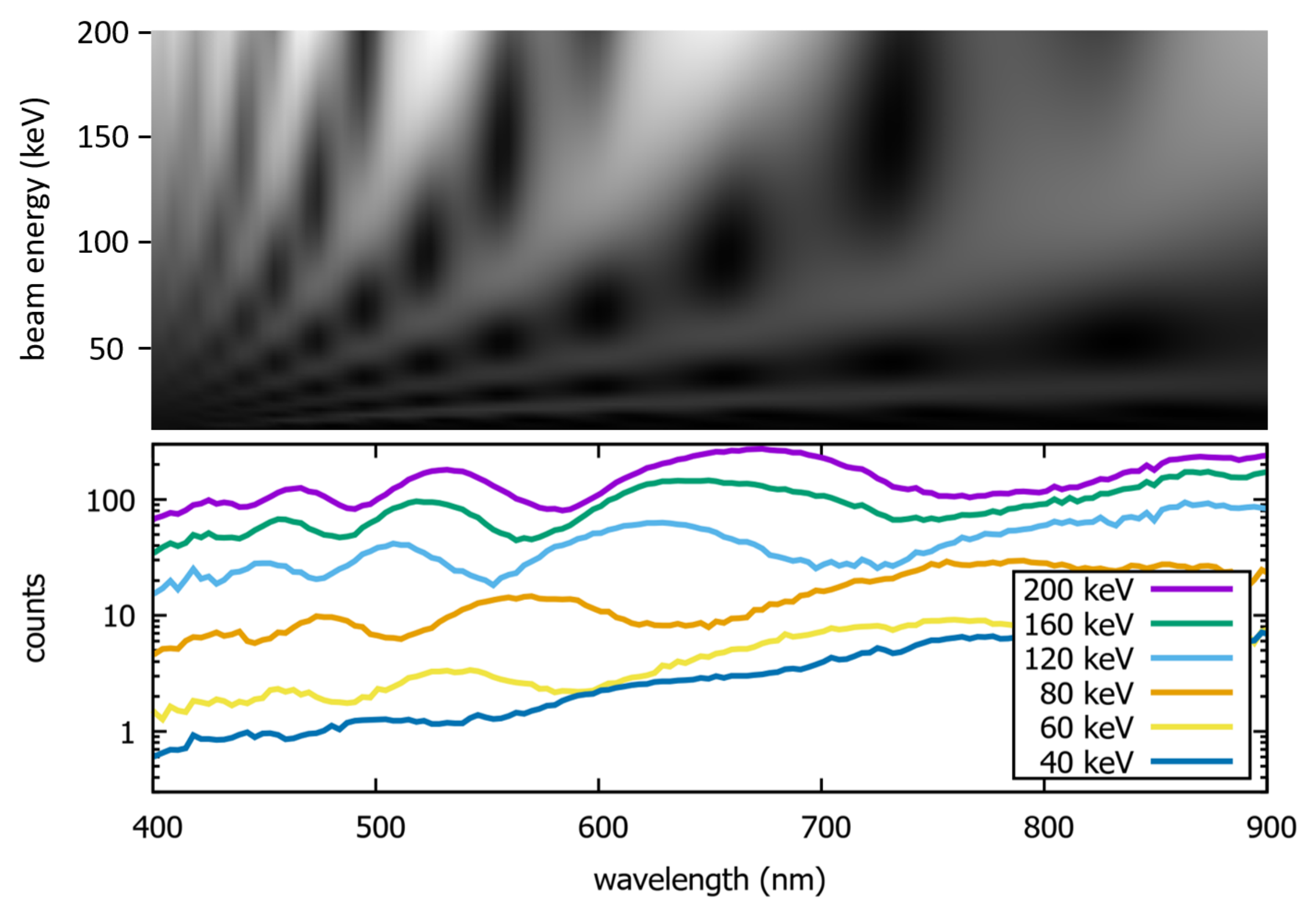}
\end{center}
\caption{Top: Simulation of the CL spectrum intensity with respect to beam energy (y-axis) and emitted wavelength (x-axis) for a sample thickness of 700~nm. Bottom: Experimental CL spectra of silicon recorded at a sample thickness of 700~nm.}
\label{HT_sim_exp}
\end{figure}

When talking about CL one first thinks about the recombination process of electron-hole pairs created when the valence electron is excited by an energy transfer from the electron probe to the shell electron. The resulting light spectrum is -- apart from the intensity -- of course independent of the initial beam energy. The same is true for the emission of Transition radiation. But the emitted \Cerenkov light is not\cite{horak2015um}. Consequently the total inner reflection causes shifts in the minima and maxima of the interference pattern. The top of Fig. \ref{HT_sim_exp} shows the calculation of the CL intensities for a beam energy range of 10~keV to 200~keV and a constant sample thickness of 700~nm.

The experiment was performed in image mode of the TEM and the electron beam was focused to a spot of a diameter of 10-15~nm. Both mirrors of the GATAN VULCAN sample holder were read out simultaneously. The beam energy of the TEM was varied from 40~keV to 200~keV. At the bottom of Fig. \ref{HT_sim_exp} shows the spectra after correction for the system response. A blue-shift of the extrema in the interference pattern can be observed when the beam energy is decreasing, being in perfect agreement with the simulations.

\subsection{varying sample thickness}

\begin{figure}[h!]
\begin{center}
\includegraphics[width=10cm]{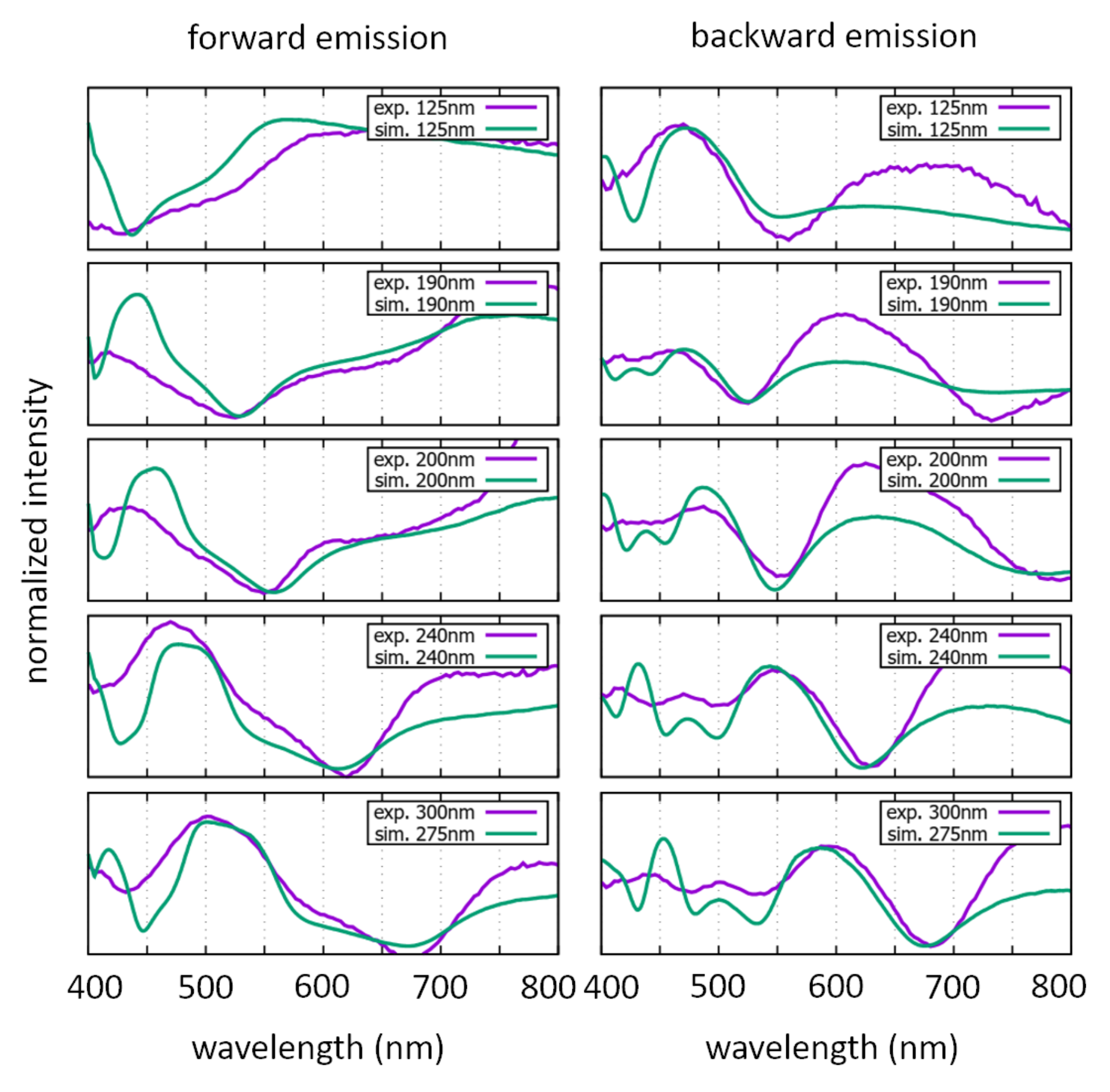}
\end{center}
\caption{Experimental (purple) and simulated (green) forward and backward emission CL spectra of silicon at sample thicknesses of 125~nm, 190~nm, 200~nm, 240~nm, and 275~nm, respectively. The experimental thickness values are determined by means of EELS. At thicknesses above 240~nm the agreement between experimentally determined thickness and simulated thickness increases. The best fit to the experimental spectral distribution of the emitted light at thicknesses of 300~nm can be found for 275~nm.}
\label{CL_back}
\end{figure}

Another parameter influencing the interference of light reflected by total inner reflection is the sample thickness $d$ in Eq. \ref{Dd}. In order to study the interference signal we utilized focused ion beam (FIB) to manufacture flat samples with thicknesses of 125~nm, 190~nm, 200~nm, 240~nm, and 275~nm, respectively. The thicknesses were determined using EELS (collection semi-angle 8.4~mrad, 200~keV electrons, image mode of the TEM), employing the log-ratio method, and by comparing the CL signal with simulations. It was found, that for thicknesses of $d/$\textit{IMFP}$< 2$ -- with \textit{IMFP} as the inelastic mean free path length -- the log-ratio method gives reasonable thickness values, whereas for larger thicknesses it gives an overestimation. Fig. \ref{CL_back} shows a comparison of the experiment (purple curve) and simulation (green curve) for forward and backward emission of the light.

For thicker sample regions the log-ratio method in EELS seem to fail. The measured CL spectrum does not fit to the simulated ones, consequently a spectrum having a better fit is chosen. The respective sample thickness of the simulation can be taken as the true sample thickness. The experimental thickness calculation in EELS is based on the fact that the intensity ratio between sample thickness and \textit{IMFP} is the natural logarithm of the ratio of the overall intensity $I_0$ to the elastically scattered electrons $I_{ZLP}$.

\begin{equation}
\frac{d}{IMFP}=\ln\frac{I_0}{I_{ZLP}}
\end{equation}

The log-ratio methods in EELS estimates 300~nm, which does not fit to the spectral distribution of the emitted light. The best fit between experiment and simulation is at 275~nm sample thickness (Fig. \ref{CL_back}). A reason for the failure of the log-ratio methods at thicker samples is, that a considerable portion of he inelastically scattered electrons can be found at higher energy losses, being not detected by the spectrometer, since the spectrometer dispersion settings allow only for a certain energy range to be recorded. Consequently the software fits a power-law up to high energy losses (1000~eV above the spectrum end). The intensity decay of the EELS spectrum has only in first approximation a power-law behavior. For an energy interval of 1000~eV a power-law fit leads to an overestimation of the total intensity $I_0$.

\begin{figure}[h!]
\begin{center}
\includegraphics[width=10cm]{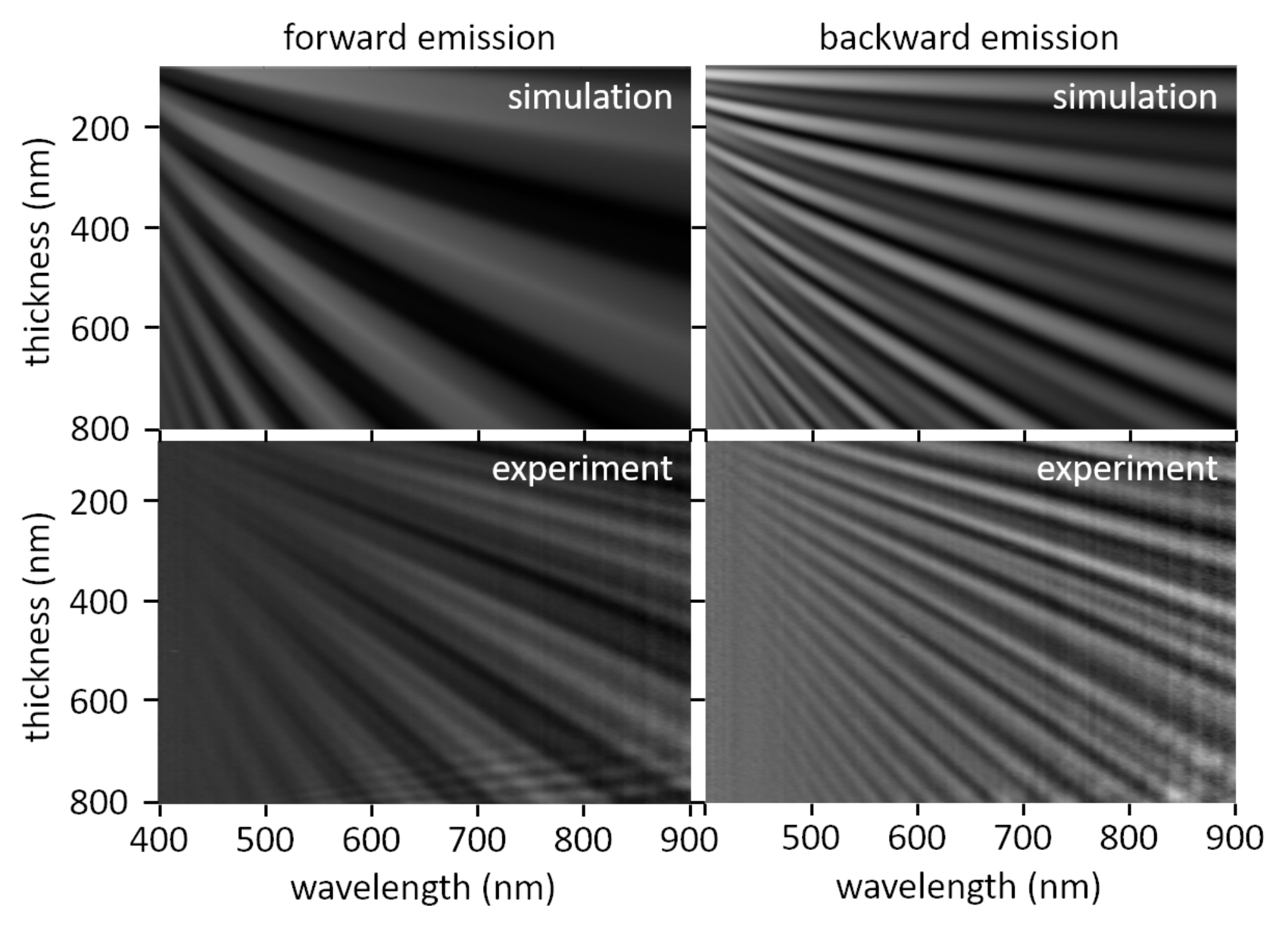}
\end{center}
\caption{Simulated and experimental forward and backward emission CL spectra of silicon for a thickness ramp between 80~nm and 800~nm. For better visibility, the intensities of all subfigures are linear but normalized such, that the darkest pixel is black and the brightest one is white.}
\label{ramp}
\end{figure}

Additionally we recorded a CL linescan across a wedge shaped sample with a thickness ramp from 80~nm to 800~nm. Figures \ref{ramp} and \ref{ramp} show the experimental results (bottom) and simulations (top) for backward and forward emission, respectively. 

The small disagreement between the simulation and the experiment is due to an imperfect thickness ramp. Anyhow an intensity modulation of the emission spectrum can be observed. A similar intensity modulation for InP was described earlier in \cite{tizei2013JPhysCondMat}. Therein it was interpreted as leaky cavity modes of and was described by a phenomenological model. In our simulations we refer to Yamamoto's theory \cite{yamamoto1996JElecMic}, which itself is based on relativistic electrodynamics.

\subsection{varying distance from sample edge}
\begin{figure}[h!]
\begin{center}
\includegraphics[width=10cm]{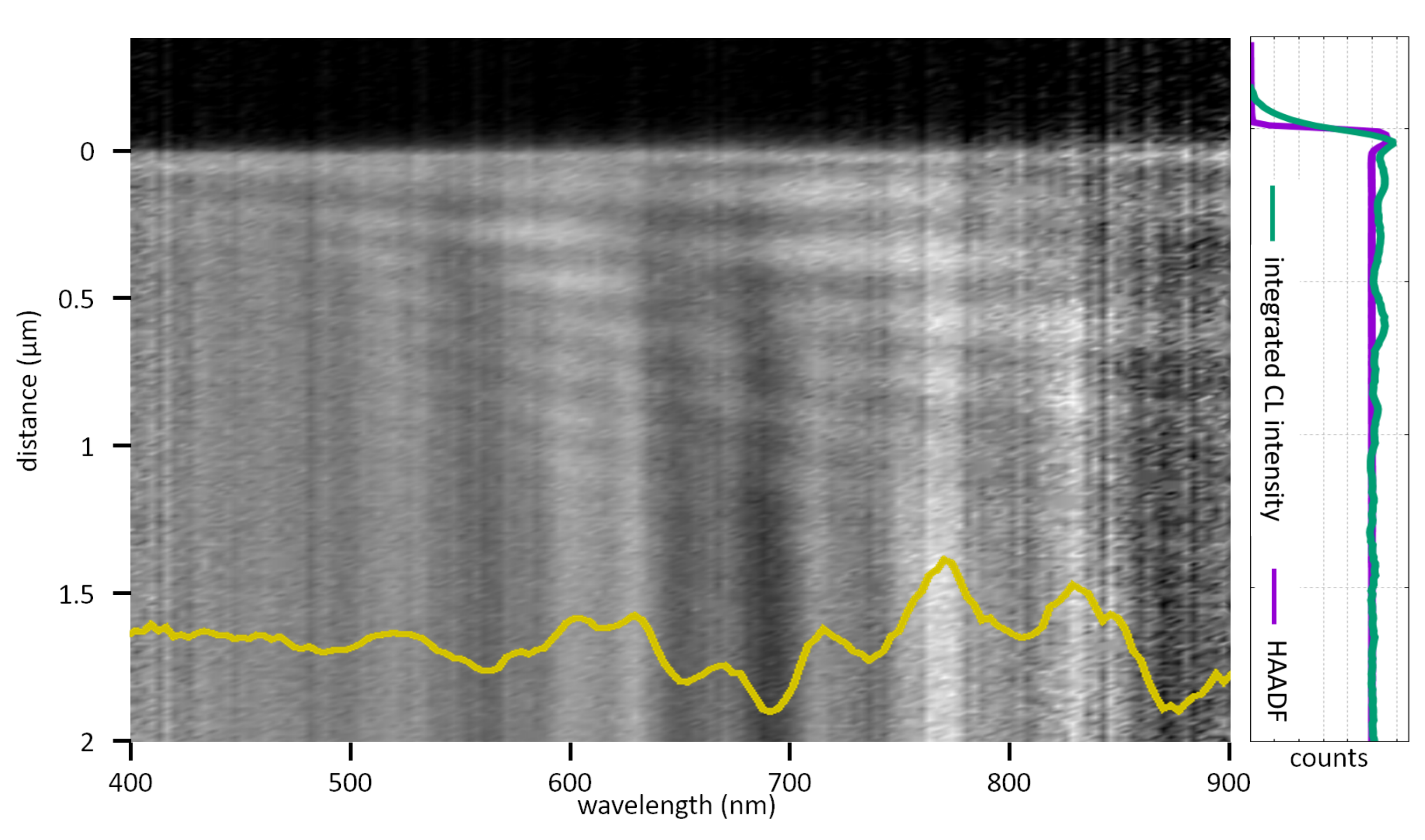}
\end{center}
\caption{Left: Experimental sum of forward and backward emission CL spectra of silicon for a linescan across the sample edge. The intensity profile is the CL spectrum recorded 2~$\mu$m away from the sample edge. Right: Experimental HAADF and integrated CL intensity profiles.}
\label{edge_exp}
\end{figure}

As already visible in the experimental parts of Figs. \ref{ramp} another interference pattern appears at the thick sample positions showing a dispersion being opposite from the main one. It is due to the vicinity of the specimens edge, which is at 1~$\mu$m distance from the end of the spectrum line scan. This kind of interference leads to a total intensity modulation, i.e. a modulation of the intensity being integrated over the whole visible range. Figure \ref{edge_exp} shows the CL-linescan across a sample edge going 2~$\mu$m deep into the constantly thick sample. The HAADF profile, being recorded simultaneously with respect to the CL signals, proofs the constant thickness. Additionally we integrated over the CL intensity and find in the CL intensity profile an intensity modulation within the first $\approx$1.5~$\mu$m. This is exactly the region, where the interference pattern from total inner reflection of the edge surface appears, which is parallel to the electron beam axis.

The yellow line in Fig. \ref{edge_exp} is the CL spectrum recorded 2~$\mu$m away from the edge. Although the sample thickness in the experiment and the calculation are not equal, the interference fringes of the sample/vacuum interface being parallel to the beam axis can be observed in both results. 

\begin{figure}[h!]
\begin{center}
\includegraphics[width=10cm]{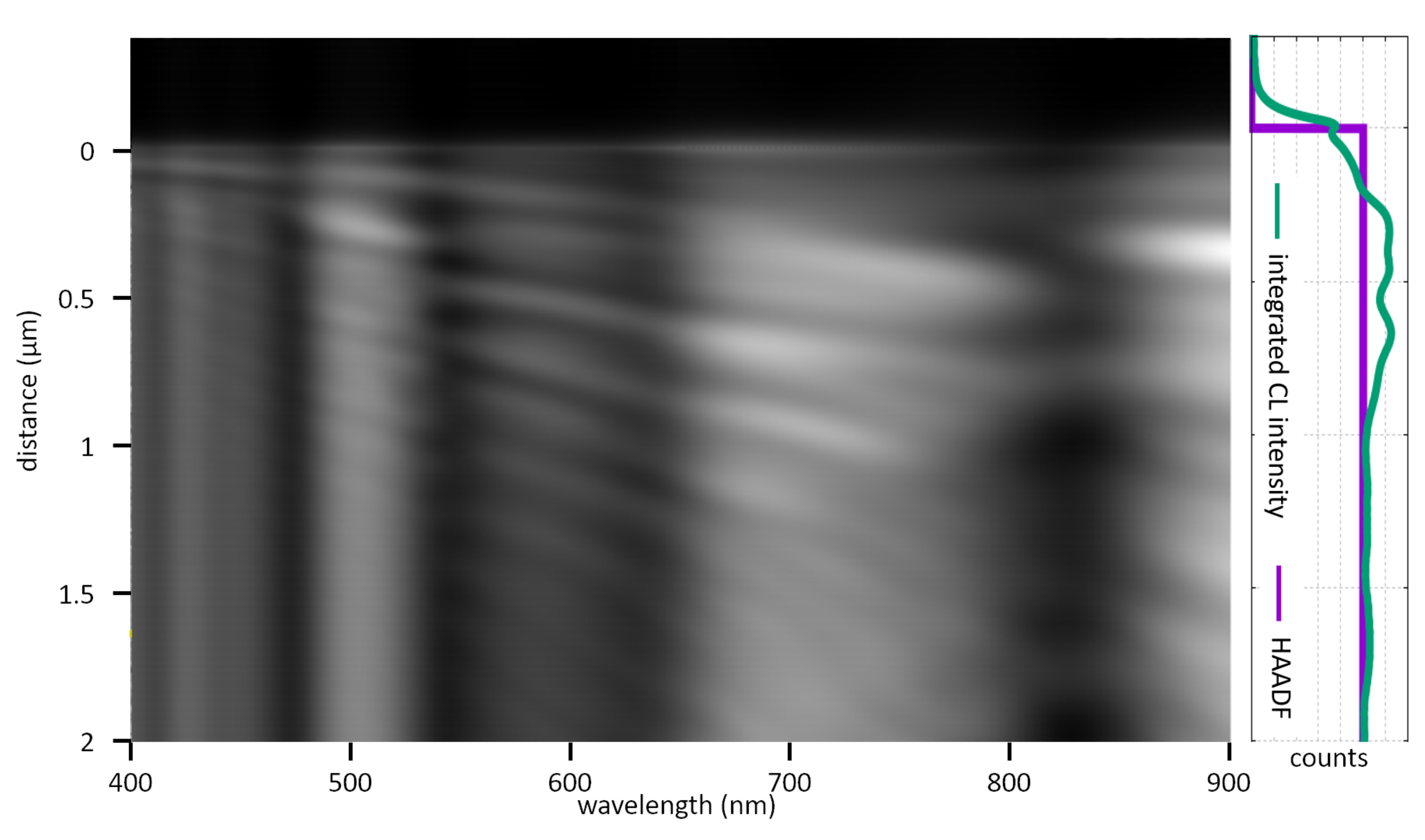}
\end{center}
\caption{Calculated sum of forward and backward emission CL spectra of silicon for a linescan across the sample edge at a sample thickness of 300~nm. The sample edge is at distance~$= 0$~$\mu$m. Right: Calculated HAADF and integrated CL intensity profiles.}
\label{edge_sim}
\end{figure}

\begin{figure}[h!]
\begin{center}
\includegraphics[width=8cm]{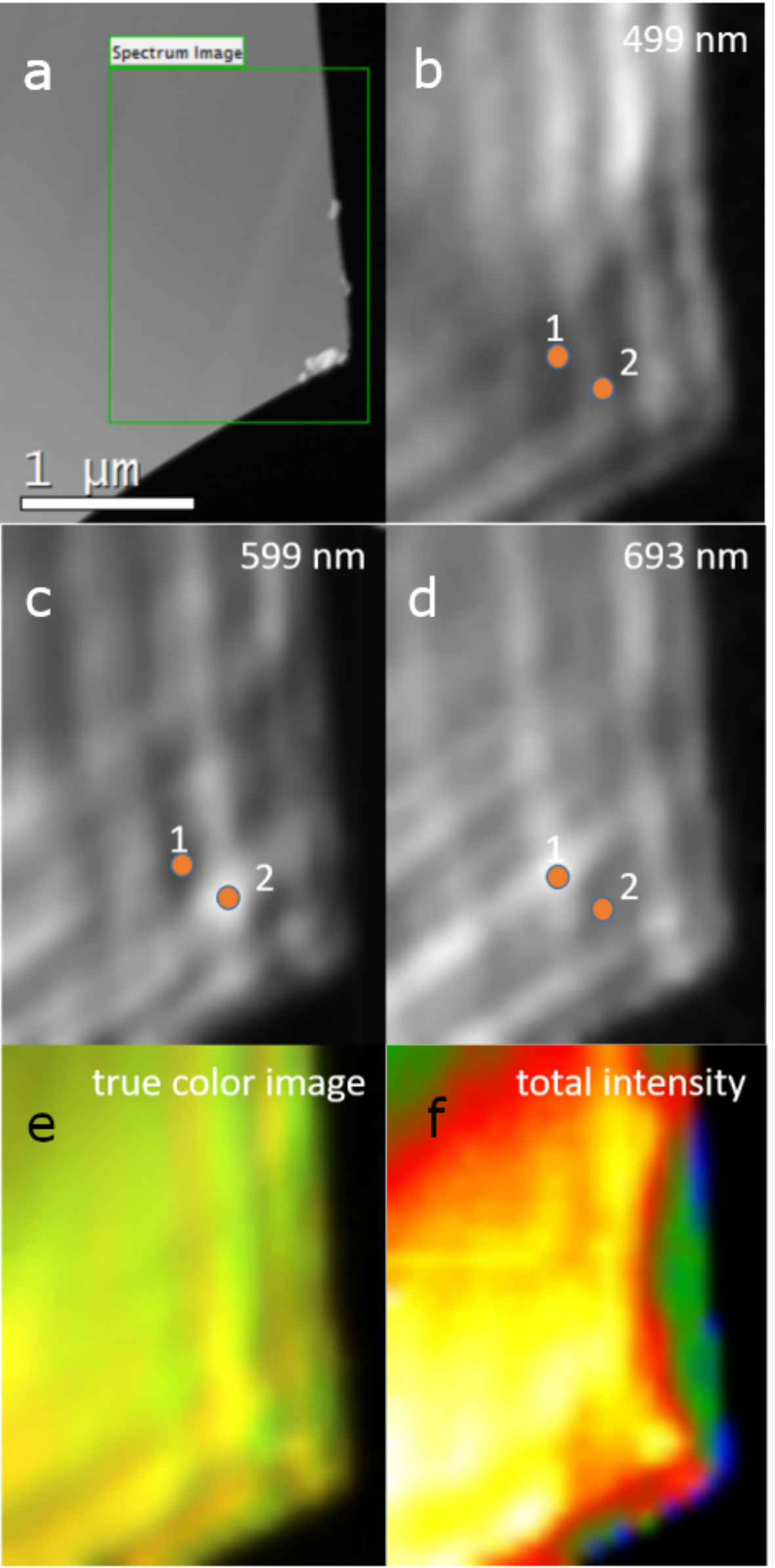}
\end{center}
\caption{a) HAADF image of the homogeneously thick Si sample with an 111$^{\circ}$ edge. b) Light emission intensity distribution of the investigated area for 499~nm wavelength. c) Light emission intensity distribution of the investigated area for 599~nm wavelength. d) Light emission intensity distribution of the investigated area for 693~nm wavelength. The spots "1" and "2" are the positions where from the spectra in Figure \ref{BiedermannSpektren} are sown. e) True color map of the emitted light. f) Normalized total CL intensity integrated over the whole visible range. Color code: rainbow scheme.}
\label{Biedermann}
\end{figure}

\begin{figure}[h!]
\begin{center}
\includegraphics[width=10cm]{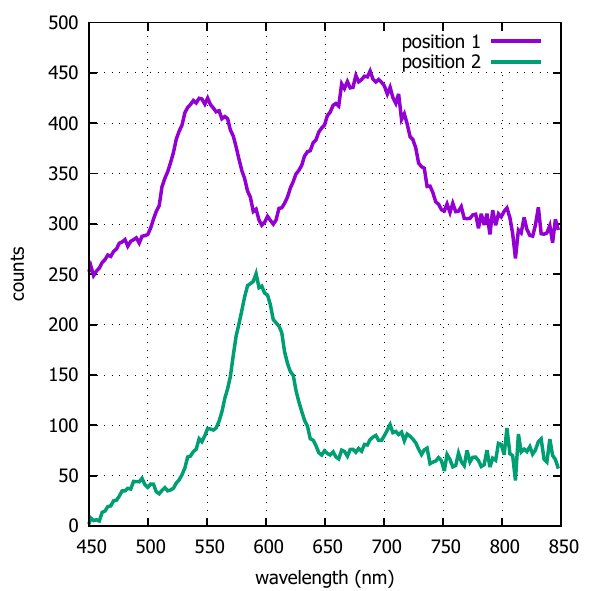}
\end{center}
\caption{CL spectra from the positions marked in Fig. \ref{Biedermann}. The spectrum from position 1 is shifted by 250 counts for better visibility.}
\label{BiedermannSpektren}
\end{figure}

If the sample has a more complicated geometry, total inner reflection at the edges as well as at the top and bottom sample surfaces can cause even more complicated interference pattern. For this purpose we prepared a 110~nm thin Si lamella of homogenous thickness showing a 110$^\circ$ edge. The green framed area in Fig. \ref{Biedermann}a) was selected for CL mapping. Looking at the spatial intensity distributions Fig. \ref{Biedermann}f)one finds modulations which are only due to the sample geometry. The two most extreme positions are presented in Fig. \ref{BiedermannSpektren}, where position 1 shows up two strong maxima at 545~nm and 693~nm, respectively. On the other hand in the spectrum taken from position 2 only a single but even brighter maximum can be found at 599~nm wavelength.

Not only the spectral shape varies, also does the total CL intensity integrated over the whole visible range. Fig. \ref{Biedermann} (bottom right) shows the respective color coded intensity plot. A modulation can be easily recognized even though the sample thickness is constant over the whole area.

\subsection{silicon nano-ribbons with various width}

\begin{figure}[h!]
\begin{center}
\includegraphics[width=10cm]{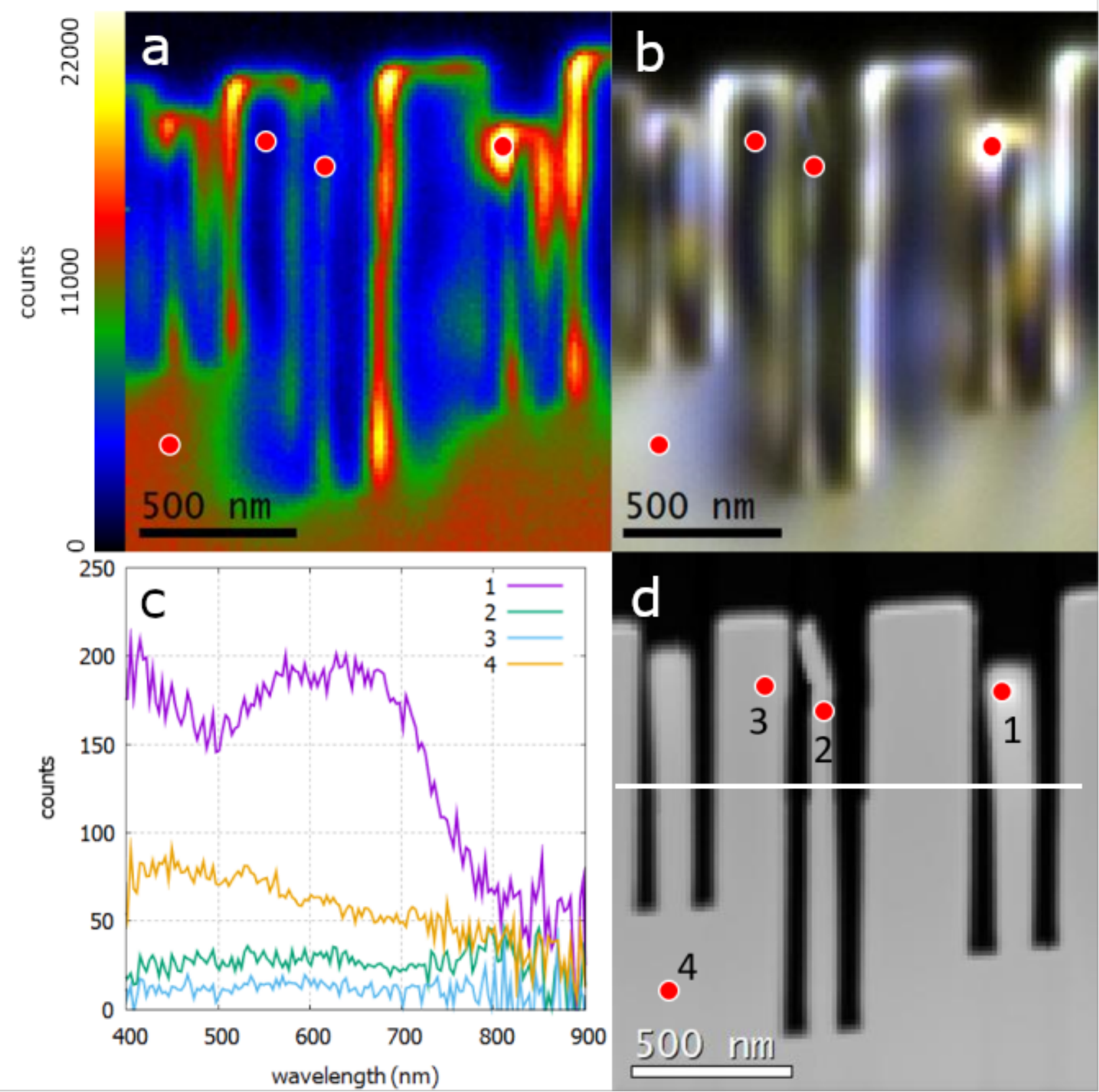}\\
\includegraphics[width=10cm]{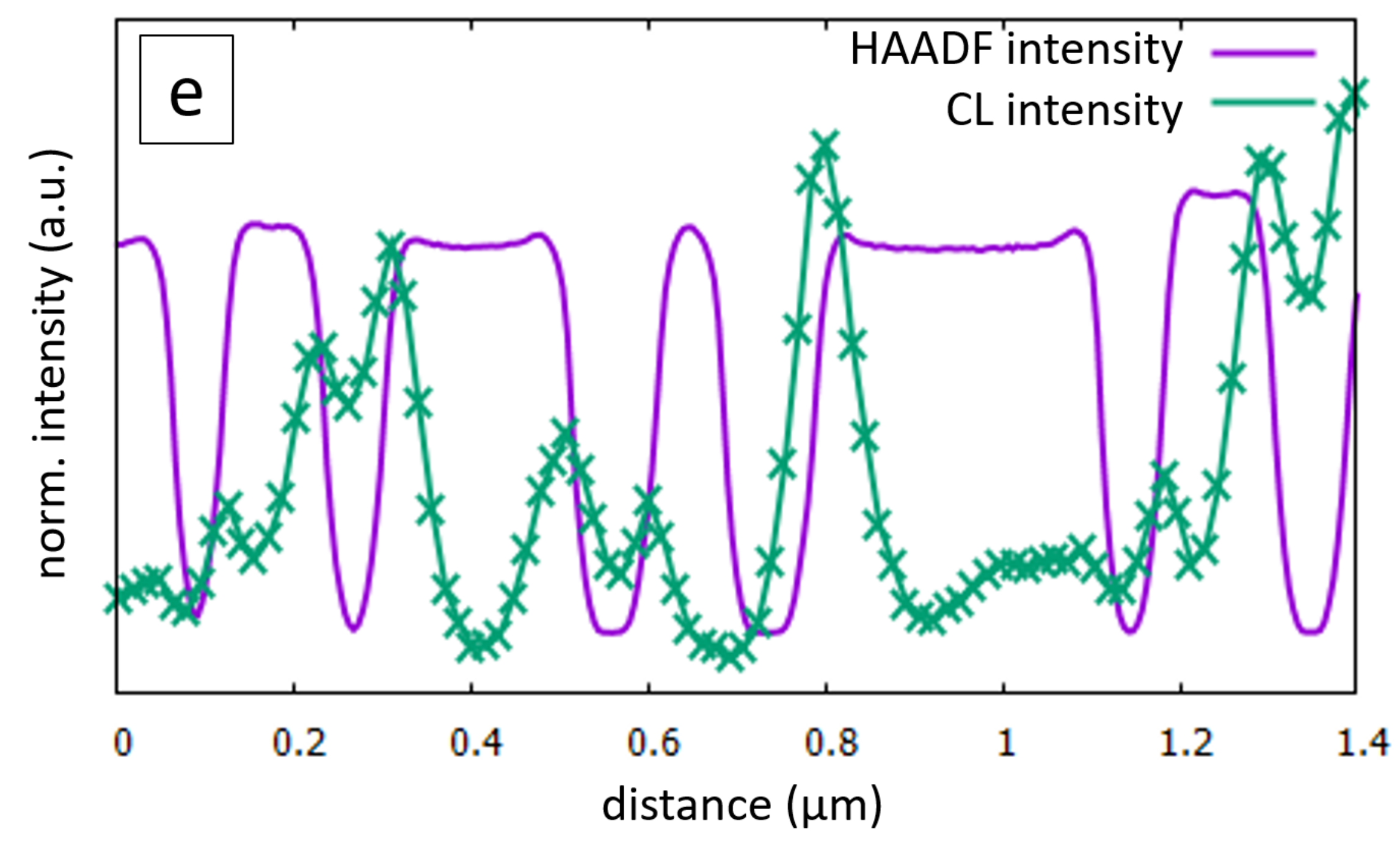}
\end{center}
\caption{a) Integrated CL intensity. b) True color image of the emitted light showing an inhomogeneity in color indicating spatially dependent spectral shapes. c) Four CL spectra taken from the indicated positions. d) HAADF image of the investigated sample area. e) Intensity profile along white line shown in (d) for the HAADF and CL intensities, respectively.}
\label{ribbons}
\end{figure}

Due to the prior generated knowledge the obvious next step is to investigate nano-structures of constantly thick Si but with different aspect ratios. For this purpose an 80~nm thin Si lamella was prepared by means of FIB. Subsequently ribbon structures were cut out of it, such that all ribbons had the same thickness of 80~nm but varying widths and lengths. CL spectrum images were recorded over such an area (Fig. \ref{ribbons}). 

When investigating FIB prepared nano-ribbons or nano-structures, several effects start to play a role. First of all there is to mention, that FIB creates a lot of point defects, thus reducing the natural luminosity of the sample. Therefore we can observe dark areas in Fig. \ref{ribbons}(b). Second, surface plasmons have to be considered, which are shape dependent. In the present situation, surface resonances are the most dominant feature in the CL spectra. Therefore the rims of the ribbons show the brightest contrast. An intensity profile across some ribbins is displayed in Fig.\ref{ribbons}(e). Third, total inner reflection on sample/vacuum-interfaces are minimized, because the width of even the broadest ribbons is too small for building up interference fringes. 

Consequently, the investigation of nano-objects, which were prepared by means of any grinding and milling technique, can lead to erroneous CL spectra, which are not only due to the nature of the sample but also due to preparation artifacts. The situation changes slightly, when investigating grown nano-structures. There is no  creation of point defects as from milling processes, thus only surface resonances and volume recombination processes will contribute to the spectral shape.

\subsection{interface between nc-Si/c-Si}

\begin{figure}[h!]
\begin{center}
\includegraphics[width=10cm]{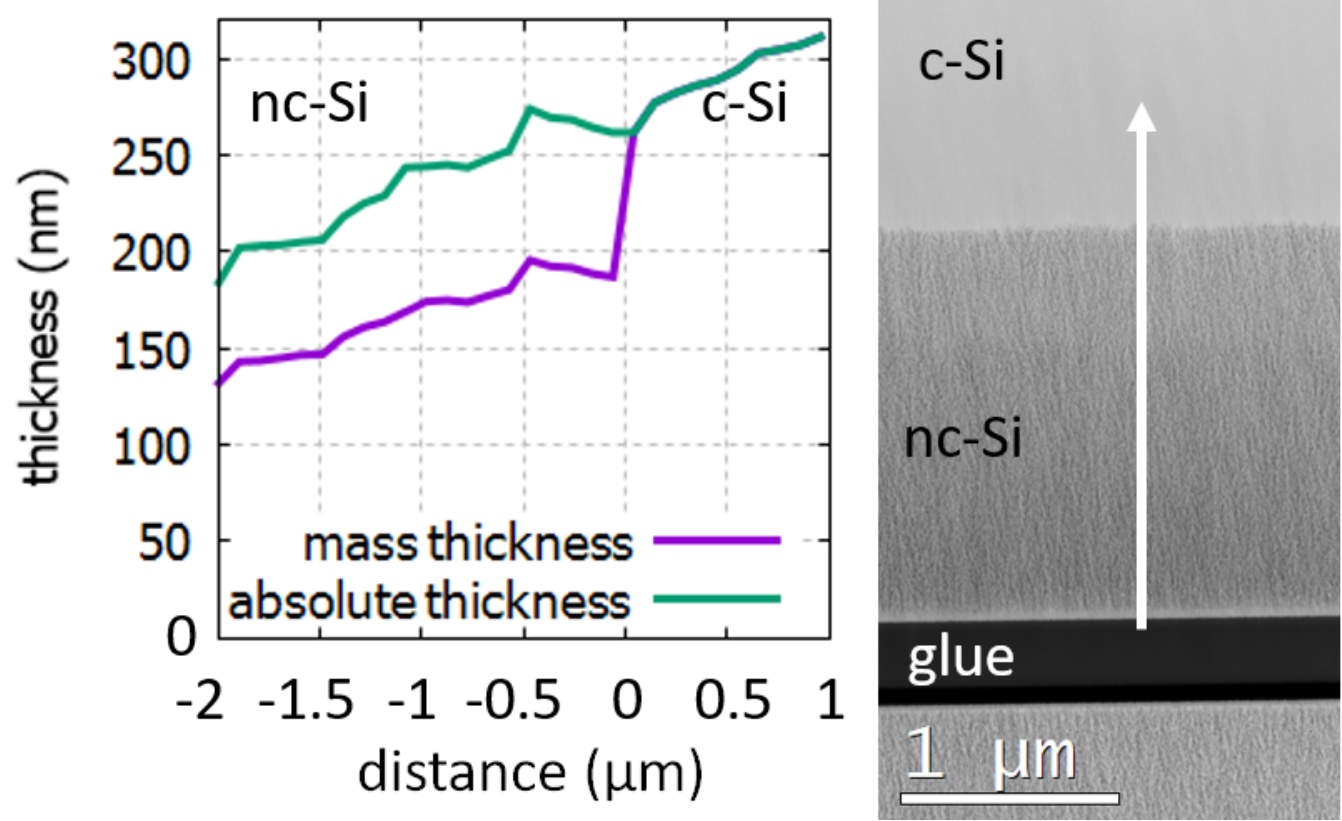}
\end{center}
\caption{Left: Thickness and mass thickness of the nc-Si/c-Si layer structure. Right: HAADF image of the investigated sample area.}
\label{fig17}
\end{figure}

In the last two subsections we discussed the contribution of total inner reflection to the CL spectrum at a sample/vacuum interface being parallel to the electron beam axis. In this subsection we focus on the influences given by a Si/Si interface. For this purpose we use a sample of nano-crystalline Si (nc-Si) deposited onto a single crystalline Si wafer (c-Si). The advantage of this material is, that the dielectric properties of both materials do not vary too much with respect to each other. On the other side, this structure has a distinct interface. Figure \ref{fig17} shows the HAADF image of the structure, with the c-Si on top, the nc-Si below and at the bottom there is the glue line of the face-to-face cross section preparation. Due to nano-pores, which appear in the nc-Si layer, the mass thickness is reduced. The mass thickness was determined via HAADF contrast and the total thickness can be calculated at the c-Si/nc-Si interface to be 140\% of the mass thickness within the nc-Si layer.

\begin{figure}[h!]
\begin{center}
\includegraphics[width=9cm]{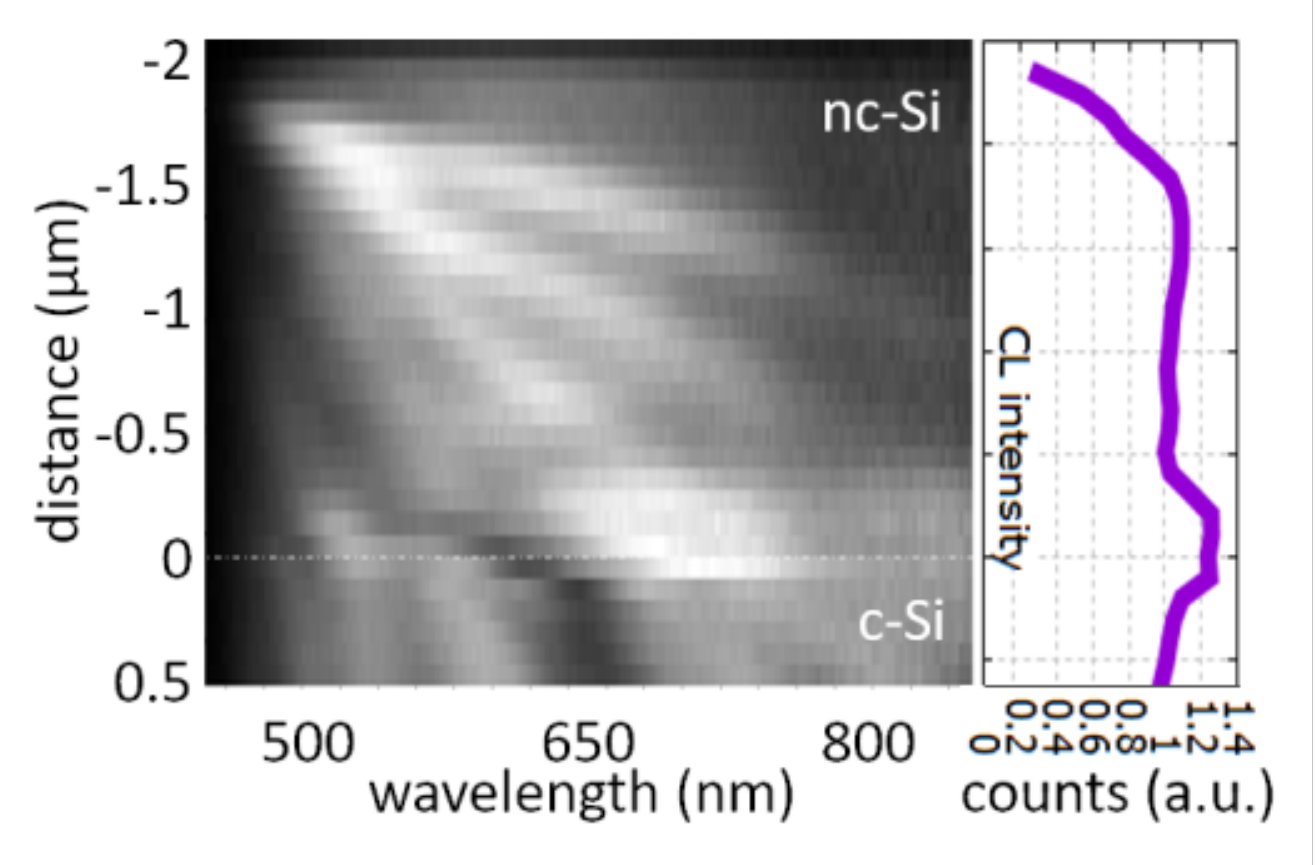}
\end{center}
\caption{Left: Oscillations in the light emission due to changes in the sample thickness. Only at the interface, a different emission spectrum is observed. Right: Integrated (400~nm - 800~nm) CL intensity with respect to the sample position. At the interface a 25\% intensity increase is observed.}
\label{fig18}
\end{figure}

\begin{figure}[h!]
\begin{center}
\includegraphics[width=9cm]{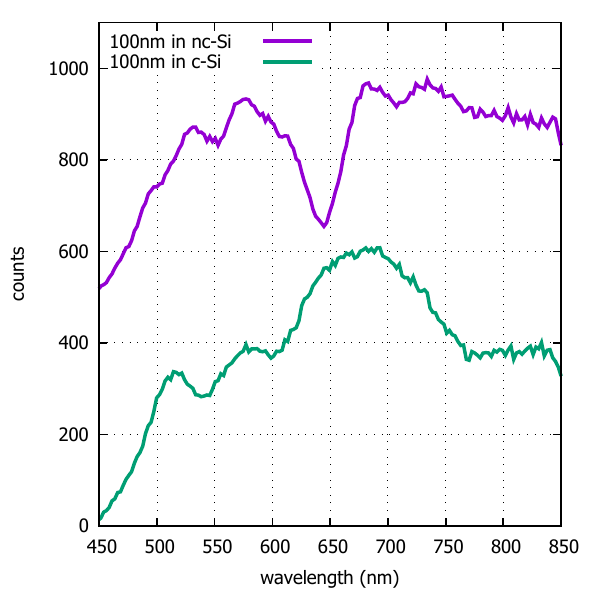}
\end{center}
\caption{Top: CL spectra from the positions 100~nm inside nc-Si and 100~nm inside c-Si, with respect to the nc-Si/c-Si interface. The spectra are shifted at the y-axis for better visibility.}
\label{fig19}
\end{figure}

Due to the fact, that the thickness shows a ramp-like behavior, red-shifting oscillations in the emission intensity can be measured. Additionally, directly at the interface (0~nm) an increase of 25~\% in the integrated CL-intensity can be observed. The oscillations are only changed within $\pm$~100~nm beside the interface. Everywhere else they obay the expected behavior with respect to the dielectric properties of silicon (see Figs. \ref{fig18} and \ref{fig19}). The spectra from nc-Si and c-Si at a distance of 100~nm with respect to the interface show again a completely different shape (Fig \ref{fig19}).

It is worth to mention, that the mass thickness is reduced by 40\% with respect to the single crystalline silicon substrate. This indicates the presence of numerous nano-pores. But contrarily to what can be expected, the numerous random interfaces in the nc-Si layer do not prevent from generation of the interference within the whole sample thickness. This can be explained by the extremely small size of the nano-crystals as well as the nano-pores being in the range of a few single nanometers.

\clearpage
\subsection{GaAs}
Up to now we were studying Si having luminescence only in the infra-red (IR) region. The direct bandgap of Si is at 3.6~eV, thus the absorption coefficient is small below that value. Moreover, the IR region cannot be probed with the used light guide of the spectrometer. Even if it would be possible, the second refraction maximum of the analysing grating inside the spectrometer would interfere with the IR spectrum of the emitted radiation. Consequently, we have chosen GaAs for testing the observed behavior on a material having a bandgap in the red region of the spectrum of light and a significant absorption coefficient above. Additionally GaAs is known to be highly luminescent, thus making it being a preferred material for high-efficient solar cell structures. It has a strong interband transition at roughly 1.4 eV. 

\begin{figure}[h!]
\begin{center}
\includegraphics[width=9cm]{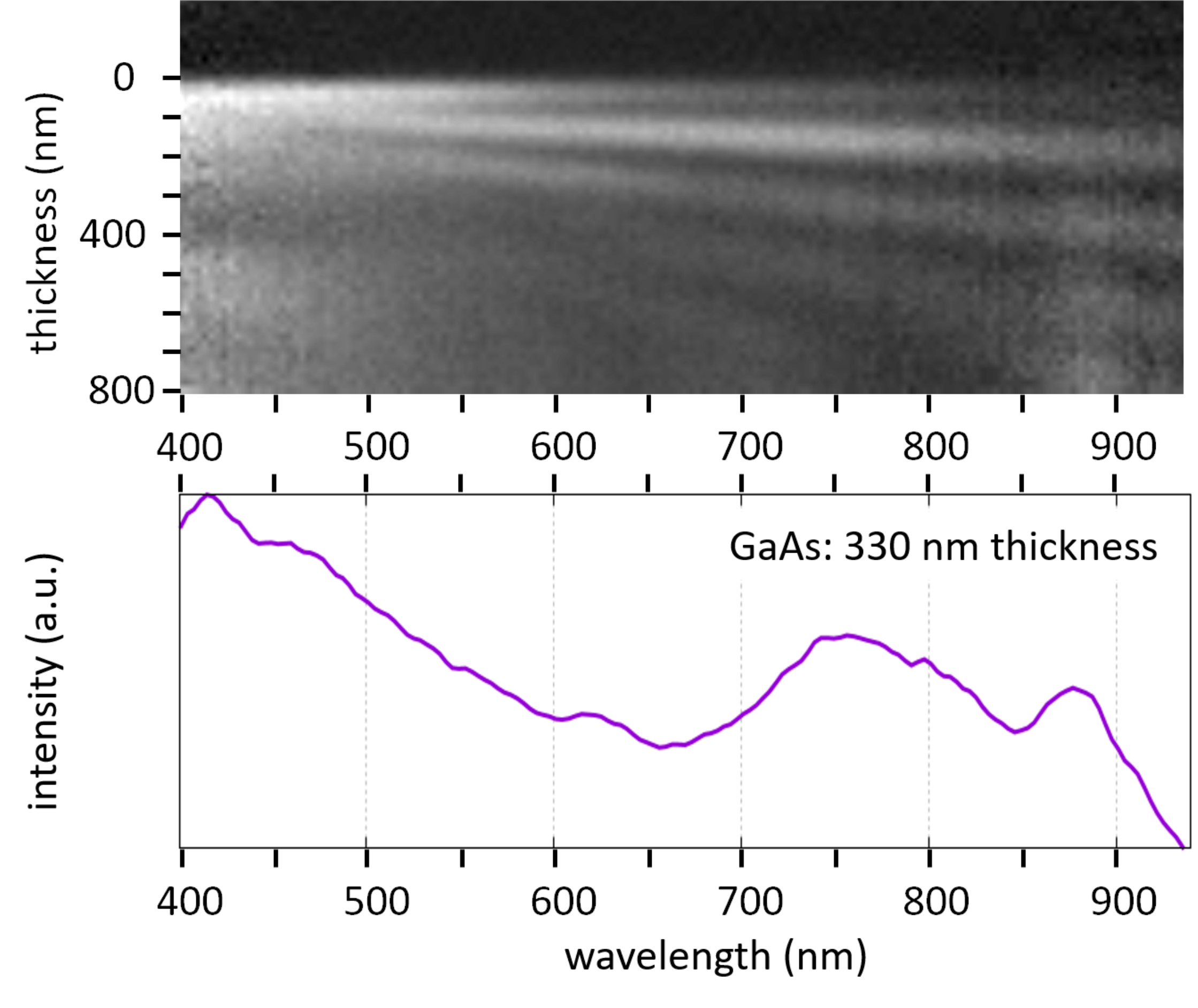}
\end{center}
\caption{CL spectra of GaAs with respect to sample thickness. The interband transition at 1.39~eV (877~nm) can be identified at all thicknesses. Bottom: GaAs CL-spectrum from the sample position at 330~nm thickness.}
\label{figGaAs}
\end{figure}

For the experiment we prepared a GaAs wedge sample by mechanical means only and without any ion beam bombardment. Thus we are sure not to create any point defects reducing the luminescence of the specimen. Figure \ref{figGaAs} shows the STEM-CL line scan perpendicular to the sample edge, such that the sample thickness increases with respect to the beam position. The thickness were measured using EELS. The interband transition at 877~nm can be identified at any sample thickness. Nevertheless, as shown as an example in the spectrum recorded at 330~nm sample thickness, the main contribution to the spectrum is due to the interference of \Cerenkov and transition radiation at the top and bottom sample surface. A similar behavior was already observed used monochromated EELS \cite{stoeger2006micron}.

\clearpage
\section{FDTD simulations}

In prior simulations, it can be seen that the agreement between experiment and simulation is good but not excellent even despite the fact that the log-ration method in EELS underestimates the real thickness of samples. A more accurate concept of calculating the emitted light is solving the Maxwell equations numerically. These numerical calculations are quite slow, even when sophisticated commercial-grade solver (Lumerical) cite{lumerical} based on a finite-difference time-domain (FDTD) method is used to perform the calculations.

In the Lumerical environment, the effects of radiation emission connected with deceleration of electrons passing through a material are modeled by a line of electric dipoles (simulating the electron beam) polarized in the direction of propagation of the electron beam. The distance between the neighboring dipoles in the line is in order of tens of nm and its temporal delay reflects the velocity of the electrons in the beam. When no structure is present, the velocity of the propagating electron remains constant and no radiation is emitted. However, in FDTD simulations, every dipole is a source of radiation regardless of the presence of any material. To solve this issue, it is necessary to run a reference simulation, with exactly the same parameters and mesh but with no material present. The electric field recorded in the reference simulation is then subtracted from the field recorded in material simulations. By this technique, the artifacts in the recorded spectra connected with the presence of emitting dipoles are eliminated. This procedure is already well established when both, EELS \cite{cao2015ACSP} and CL \cite{chaturvedi2009ACSN}\cite{das2012JPCC}, are simulated by an FDTD technique. Nevertheless, this method does not provide absolute photon emission probabilities, as the emission spectra are obtained by integrating Pointing vector propagating through an area of arbitrary size above or below the material. Therefore the results obtained by this method are normalized to one. 

\begin{figure}[h!]
\begin{center}
\includegraphics[width=10cm]{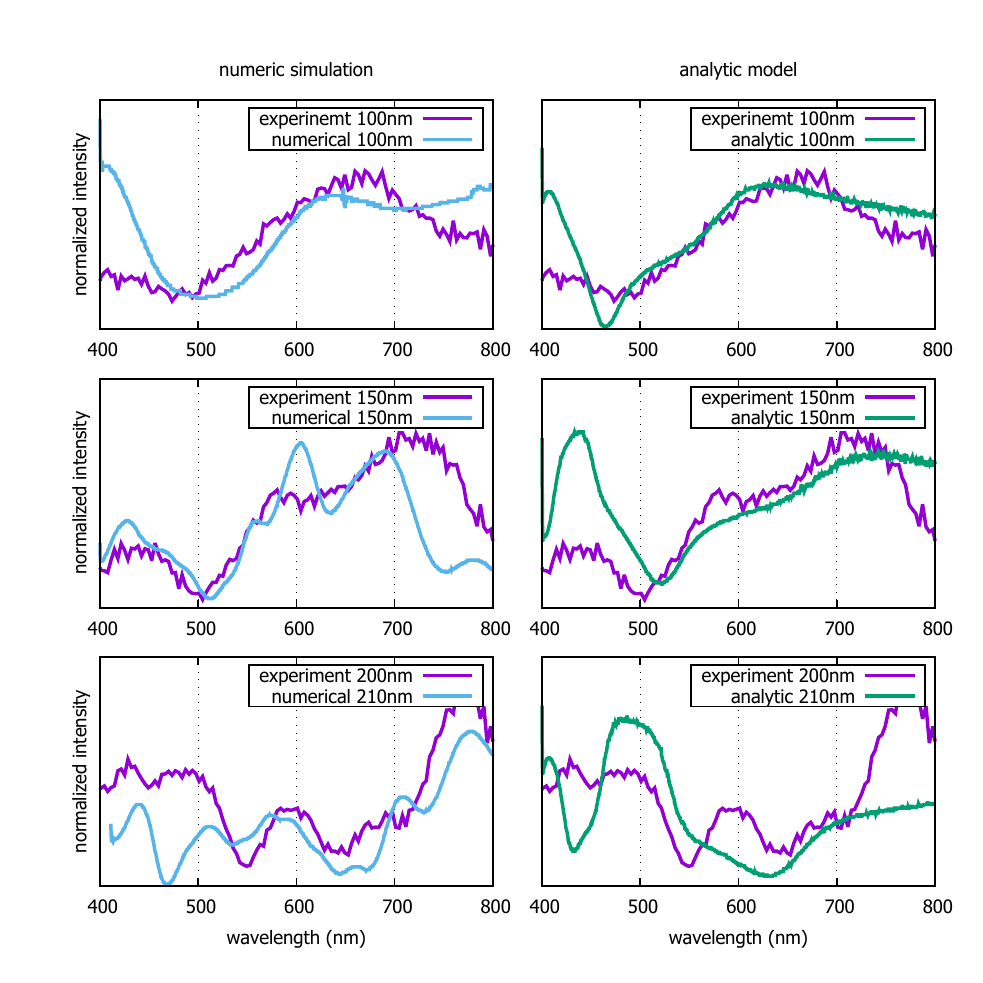}
\end{center}
\caption{Comparison of CL spectra simulated anaytically (theory), numerically (lumerical) and the measured ones (experiment), respectively.}
\label{FDTD}
\end{figure}

For the present simulations we assumed a wedge-shaped specimen, as prepared by FIB. Comparison with the experimentally obtained results gives a better agreement than the analytic simulation does. Fig. \ref{FDTD} shows an excerpt of spectra from the numerical simulation (left column, blue spectra), the experiment (purple spectra) and the analytic simulation (right column, green spectra).

\section{Discussion}

\begin{figure}[h!]
\begin{center}
\includegraphics[width=10cm]{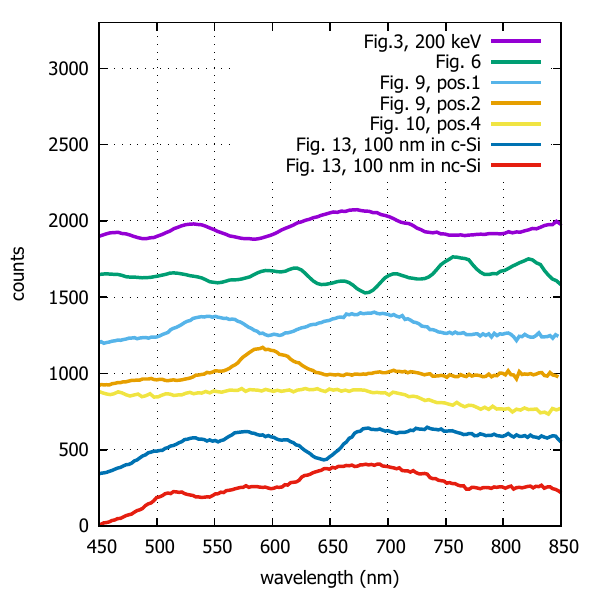}
\end{center}
\caption{Comparison of some CL spectra from throughout this manuscript. The spectra are shifted at the y-axis for better visibility.}
\label{discussion}
\end{figure}

In the present paper we have investigated the luminescence of silicon and GaAs. We found that the CL spectrum is dominated by interference caused by the total inner reflection of all interfaces between the sample and vacuum or interfaces inside the sample for both materials. There is no doubt, that excitonic emissions in semiconductors can be measured by employing CL. CL is also straight forward for plasmonic investigation on (metallic) nano-objects. But when semiconductors shall be investigated, directly interpretable CL spectra cannot be recorded. A comparison to simulation including the excitation of Cerenkov radiation and transition radiation is indispensable. Fig. \ref{discussion} gives a comparison of some CL spectra being presented within this manuscript. All of them are from undoped silicon. Nevertheless, it is clear immediately that a direct interpretation is impossible due to the varying spectral shapes caused by interference fringes.

In a scanning electron microscope (SEM) the situation is completely different, because the sample geometry shows a nearly infinite thickness (in relation to the wave length of visible light). But backscattered electrons may also excite \Cerenkov radiation which is then collected by a CL system, too. 

\section*{Acknowledgements}

Part of this work was supported by MEYS CR under the projects CEITEC Nano Research Infrastructure (project No. LM2015041, 2016-2019) and CEITEC 2020 (project No. LQ1601). This research was in part supported by the SINNCE project of the European Union's Horizon 2020 programme under the grant agreement No. 810626.



\begin{thebibliography}{10}
\expandafter\ifx\csname url\endcsname\relax
  \def\url#1{\texttt{#1}}\fi
\expandafter\ifx\csname urlprefix\endcsname\relax\def\urlprefix{URL }\fi

\bibitem{kociak2014a}
M.~Kociak, O.~Stephane, Mapping plasmons at the nanometer scale in an electron
  microscope, Chem Soc Rev 43 (2014) 3865--3883.

\bibitem{goetsch2018PRM}
T.~Götsch, E.~Bertel, A.~Menzel, M.~Stöger-Pollach, S.~Penner, Spectroscopic
  investigation of the electronic structure of Yttria-stabilized Zirconia,
  Phys. Rev. Materials 2 (2018) 035801.

\bibitem{goetsch2017ECSt}
T.~Götsch, A.~Menzel, E.~Bertel, M.~Stöger-Pollach, S.~Penner, The
  crystallographic and electronic phase diagrams of Yttria-stabilized Zirconia
  model electrolytes, ECS Transactions 78 (2017) 311--319.

\bibitem{miller2017SensAcB}
D.~R. Miller, R.~E. Williams, S.~A. Akbar, P.~A. Morris, D.~McComb,
  STEM-cathodoluminescence of SnO$_2$ nanowires and powders, Sensors and Actuators
  B 240 (2017) 193--203.

\bibitem{stoeger2008micron}
M.~St{\"o}ger-Pollach, Optical properties and band gaps from low loss EELS:
  pitfalls and solutions, Micron 39 (2008) 1092--1110.

\bibitem{kociak2014b}
M.~Kociak, O.~Stephan, A.~Gloter, L.~F. Zagonel, L.~H.~G. Tizei, M.~Tence,
  K.~March, J.~B. Blazit, Z.~Mahfoud, A.~Losquin, S.~Meuret, C.~Colliex, Seeing
  and measuring in colors: Electron microscopy and spectroscopies applied to
  nano-optics, Comptes Rendus Physique 15 (2014) 158--175.

\bibitem{frank1966SovPhysUsp}
I.~M. Frank, Transition radiation and optical properties of matter, Sov. Phys.
  Uspekhi 8 (1960) 729--742.

\bibitem{stoeger2017um_a}
M.~Stöger-Pollach, L.~Kachtik, B.~Miesenberger, P.~Retzl, Transition radiation
  in EELS and cathodoluminescence, Ultramicroscopy 173 (2017) 31--35.

\bibitem{cerenkov1934DAN}
P.~A. \Cerenkov, Visible emission of clean liquids by action of $\gamma$
  radiation, Dokl. Akad. Nauk. SSSR 2 (1934) 451.

\bibitem{stoeger2006micron}
M.~St{\"o}ger-Pollach, H.~Franco, P.~Schattschneider, S.~Lazar, B.~Schaffer,
  W.~Grogger, Z.~H. W, \Cerenkov losses: a limit for bandgap determination and
  Kramers-Kronig Analysis, Micron 37(5) (2006) 396--402.

\bibitem{horak2015um}
M.~Horak, M.~St{\"o}ger-Pollach, The \Cerenkov limit of Si, GaAs and GaP in electron energy loss spectrometry, 
  Ultramicroscopy 157 (2015) 73--78.

\bibitem{yamamoto2001}
N.~Yamamoto, K.~Araya, A.~Toda, H.~Sugiyama, Light emission from surfaces, thin
  films and particles induced by high-energy electron beam, Surf. Int. Analysis
  31 (2001) 79--86.

\bibitem{yamamoto1996JElecMic}
N.~Yamamoto, A.~Toda, K.~Araya, Imaging of transition radiation from thin films
  on a silicon substrate using a light detection system combined with TEM, J.
  Electron Microsc. 45 (1996) 64--72.

\bibitem{termikaelian1972}
M.~Ter-Mikaelian, High-energy electromagnetic processes in condensed media,
  Interscience tracts on physics and astronomy, Wiley-Interscience, 1972.

\bibitem{ginzburg1946}
V.~L.~Ginzburg, I.~M.~Frank, JETP (USSR) 16 (1946), 15--27

\bibitem{green2008SEMSC}
M.~A. Green, Self-consistent optical parameters of intrinsic silicon at 300 K
  including temperature coefficients, Solar Energy Materials and Solar Cells 92
  (2008) 1305--1310.

\bibitem{tizei2013JPhysCondMat}
L.~H.~G. Tizei, L.~F. Zagonel, M.~Tence, O.~Stephan, M.~Kociak, T.~Chiaramonte,
  D.~Ugarte, M.~A. Cotte, Spatial modulation of above-the-gap
  cathodoluminescence in InP nanowires, J. Phys: Cond. Matt. 25 (2013) 505303
  1--6.

\bibitem{lumerical}
FDTD solutions in Lumerical, Lumerical Inc., Vancouver, CAN.

\bibitem{cao2015ACSP}
Y.~Cao, A.~Manjavacas, N.~Large, P.~Nordlander, Electron Energy-Loss Spectroscopy Calculation in Finite-Difference Time-Domain Package, ACS Photonics 2 (2015), 369--375. 

\bibitem{chaturvedi2009ACSN}
P.~Chaturvedi, K.~H.~Hsu, A.~Kumar, K.~H.~Fung, J.~C.~Mabon, N.~X.~Fang, Imaging of Plasmonic Modes of Silver Nanoparticles Using High-Resolution Cathodoluminescence Spectroscopy, ACS Nano 3 (2009), 2965--2974.

\bibitem{das2012JPCC}
P.~Das, T.~K.~Chini, J.~Pond, Probing Higher Order Surface Plasmon Modes on Individual Truncated Tetrahedral Gold Nanoparticle Using Cathodoluminescence Imaging and Spectroscopy Combined with FDTD Simulations, J. Phys. Chem. C 116 (2012), 15610--15619.

\end{thebibliography}

\end{document}